\documentclass[12pt]{iopart}

\usepackage{graphicx, subfigure}
\usepackage{SIunits}

\usepackage{setspace}
\begin{document}

\title[Enhanced THz extinction of single plasmonic antennas with conically tapered waveguides]{Enhanced THz extinction of single plasmonic antennas with conically tapered waveguides}
\author{M C Schaafsma$^1$, H Starmans$^{1,2}$, A Berrier$^1$ and J G\'{o}mez Rivas$^{1,3}$}
\address{$^1$ Center for Nanophotonics, FOM Institute AMOLF, c/o Philips Research Laboratories, High Tech Campus 4, 5656 AE Eindhoven, The Netherlands}
\address{$^2$ Optics Research Group, Faculty of Applied Sciences, Delft University of Technology, Lorentzweg 1, 2628 CJ Delft, The Netherlands}
\address{$^3$ COBRA Research Institute, Eindhoven University of Technology, P.O. Box 513, 5600 MB Eindhoven, The Netherlands}
\ead{m.schaafsma@amolf.nl}

\begin{abstract}
We demonstrate experimentally the resonant extinction of THz radiation by a single
plasmonic bowtie antenna, formed by two n-doped Si monomers with a
triangular shape and facing apexes. This demonstration is achieved
by placing the antenna at the output aperture of a conically tapered waveguide, which enhances the intensity of the incident THz field at the antenna position by a factor 10. The waveguide also suppresses the background radiation that otherwise is
transmitted without being scattered by the antenna. Bowtie antennas,
supporting localized surface plasmon polaritons, are relevant
due to their ability of resonantly enhancing the field
intensity at the gap separating the two triangular elements. This
gap has subwavelength dimensions, which allows the concentration of THz radiation beyond the diffraction limit. The combination of
a bowtie plasmonic antenna and a conical waveguide may serve as a platform for far-field THz time-domain spectroscopy of single nanostructures placed in the gap.
\end{abstract}

\pacs{71.45.Gm, 41.20.Jb, 84.40.-x, 84.40.Az}

\submitto{\NJP}

\maketitle


\section{Introduction}
Terahertz time-domain spectroscopy (THz-TDS) has a great potential for the
investigation of fundamental transitions in organic and inorganic
molecules and nanostructures. Translational and rotational degrees
of freedom in polyatomic gases and bio-molecules, lattice vibrations
in crystalline structures, conduction electrons in metals and
semiconductors, all have resonances at THz frequencies in the range of
\unit{0.1-3}{\tera\hertz}~\cite{ferguson2002,tonouchi2007,dexheimer2008,jepsen2010}. In
typical far-field implementations of THz spectroscopy, the
wavelength greatly exceeds the length scale of the individual
objects under investigation, e.g., bio-molecules or nanostructures.
Therefore, spectroscopy of single objects is challenging and
measurements are usually performed in large ensembles at high
concentrations. When working with a limited amount of material or
individual objects, the response may drop below the detection threshold.
In order to compensate for these constraints, local field
enhancements into subwavelength volumes are needed. Resonant
conducting structures, sustaining localized surface plasmon
resonances (LSPRs), are key in realizing these large local field
enhancements \cite{shi2009,tian2010,berrier2010,giannini2010}.

In this manuscript, we demonstrate that it is possible to detect and
measure the extinction of a single bowtie antenna in a standard THz
time-domain spectrometer. The antenna is formed by two triangular
monomers with facing apexes and separated by a 5 micron gap. The
monomers are made of doped Si with a metallic behavior at THz
frequencies \cite{saxler2004}. The investigated antenna exhibits an
LSPR at 0.4 THz when excited by a plane wave polarized along its
long axis. This resonance is the result of the coherent oscillation
of the free charge carriers, harmonically driven by the incident THz
electric field. Plasmonic bowtie antennas can locally enhance the
field at the LSPR frequency by several orders of magnitude in the
gap between the individual monomers. This gap has typically a volume
of $\sim 10^{-6} \lambda^3$. The large field enhancement in deep
subwavelength volumes opens the possibility for THz spectroscopy of
single nanostructures or at very low concentrations of material.
However, the scattering and extinction cross sections of such a
plasmonic antenna can be still small compared to the beam size of
the THz pulse in standard THz-TDS setups. To make the detection of a
single antenna possible we use a conically tapered waveguide.
Tapered waveguides have been proposed to guide and enhance THz
radiation into a confined region~\cite{rusina2008,kim2010,zon2011}.
Zhang {\emph et al.} have investigated the adiabatic compression of
THz radiation in tapered parallel-plate waveguides \cite{zhang2005}.
Zhan {\emph et al.} have shown superfocussing of THz radiation using
tapered parallel-plate waveguides \cite{zhan2010}. W\"achter {\emph
et al.} have used tapered photoconductive THz field probes for
subwavelength imaging \cite{wachter2009}. Enhanced THz transmission
through conically tapered waveguides has been recently reported by
Nguyen {\rm et al.}~\cite{nguyen2010}. We extend here the
application domain of conically tapered waveguides by measuring the
extinction of single resonant antennas located at the waveguide
output, where the field is enhanced.

\begin{figure}%
\centering
\includegraphics[width=12 cm]{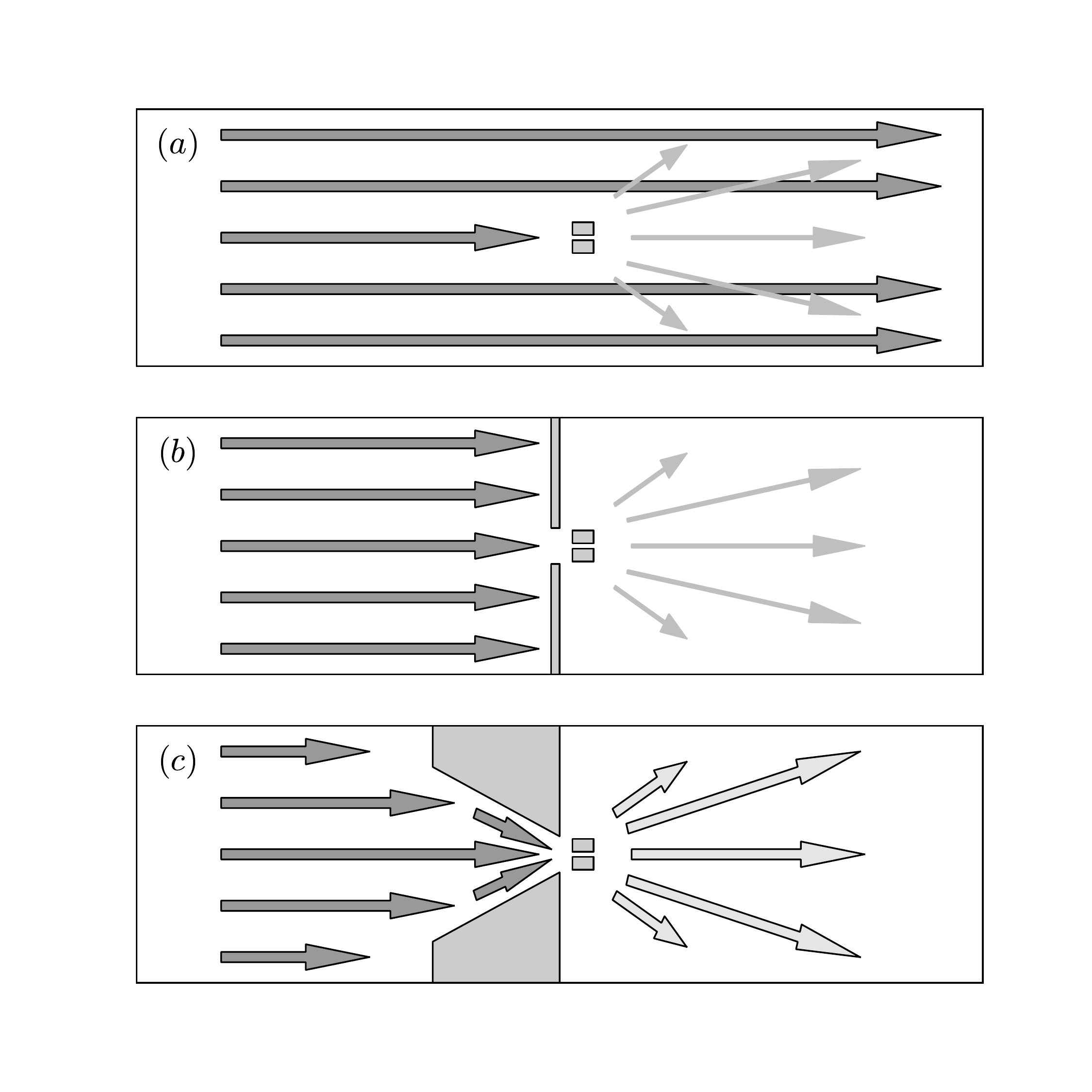}%
\caption{Schematic representation of experiments conducted on a single antenna. (a) A
plasmonic antenna with a small cross section is illuminated by a
spatially extended beam. Only a small fraction of the field is
scattered and absorbed by the antenna. In a transmission experiment, there is a
large background of unscattered radiation. (b) A cylindrical
waveguide or a metallic screen will block this background. However,
the incident intensity is also reduced. (c) The enhancement of this intensity is possible
by concentrating the incident field on the antenna with a conically tapered waveguide.}%
\label{Fig01}%
\end{figure}

A schematic of an experiment conducted on a single antenna is shown
in figure \ref{Fig01}(a). A linearly polarized THz beam is used to
illuminate a single plasmonic antenna. Since the antenna has an
extinction cross section much smaller than the size of the beam,
only a small fraction of the incident field is extinct by scattering
and absorption in the antenna. The majority of the field is
transmitted as unperturbed by the antenna, leading to a large
background in extinction measurements. In order to reduce this
background, a thin metallic screen can be used to only transmit the relevant part
of the THz beam that interacts with the antenna (figure
\ref{Fig01}(b)). Although the signal-to-background ratio is improved
in this configuration, only a small fraction of the pulse is used
and the signal-to-noise is not increased. To enhance the
signal-to-noise, it is possible to use a tapered waveguide. The
conical design guides the off-center parts of the field in the
direction of the antenna, resulting in an enhanced THz
electromagnetic field at the output aperture. This field enhancement
leads also to the enhanced response of the antenna, allowing a
more sensitive spectroscopy.

The article is organized as follows: The fabrication of the bowtie
antenna and the waveguides is described in section~\ref{Fab}. In
section~\ref{Sim}, we present simulations of the near field
enhancement of a single bowtie antenna and the conical waveguide, as
well as its enhanced transmission. THz extinction measurements of a
THz antenna at the output of the conical waveguide are compared with
the extinction of a random array of similar antennas in
section~\ref{Exp}. These measurements demonstrate the enhanced
extinction of single antennas mediated by the enhanced field at the
output of the waveguide. The article is ended with the conclusions.

\section{Waveguide and antenna fabrication}\label{Fab}
\begin{figure}%
\centering
\includegraphics[width=12 cm]{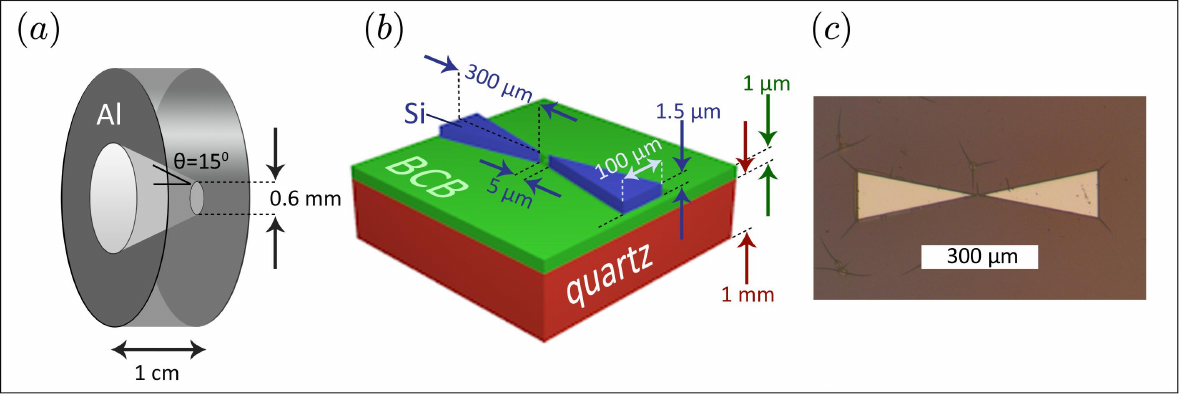}%
\caption{Schematic representation of (a) the conical waveguide and
(b) the bowtie antenna. (c) Optical microscope image of the bowtie
antenna. The antenna is formed by two triangular monomers  of
n-doped Si, bonded onto a quartz substrate with BCB. The cracks in
the microscope
image are in the supporting BCB layer.}%
\label{Fig02}%
\end{figure}

Two waveguides have been fabricated for the experiments. A conical
shape waveguide is used to enhance the field at its output aperture and a
cylindrical waveguide with the same output dimensions is used to
reference the transmittance. The waveguides have been fabricated using
electrical discharge machining: A conically tapered electrode with a
half-angle of $15^o$ or a cylindrical electrode with diameter of
\unit{0.6}{\milli\meter} are lowered through an aluminium plate,
eroding a conical or cylindrical hole. The dimensions of the
conically tapered waveguide are a thickness of
\unit{1}{\centi\meter}, an output aperture with a diameter of \unit{0.6}{\milli\meter}
and a half-angle of $15^o$. The thickness of the cylindrical
waveguide is \unit{0.5}{\milli\meter} and its diameter is
\unit{0.6}{\milli\meter}. A schematic representation of the conical
waveguide is shown in figure \ref{Fig02}(a).

The semiconductor plasmonic antennas have been fabricated using
conventional micro-fabrication techniques~\cite{berrier2012}: A
silicon-on-insulator (SOI) wafer with a \unit{1.5}{\micro\meter}
thick undoped top layer is implanted with arsenic atoms, introducing
a free carrier concentration of $(6 \pm 3)\times
\unit{10^{19}}{\centi\meter^{-3}}$. This doping level makes silicon
a conductor at terahertz frequencies. The SOI wafer is bonded onto a
\unit{1}{\milli\meter} thick quartz substrate with benzocyclobutene
(BCB). The silicon substrate and oxide layer are subsequentially
removed with wet chemical etching using KOH and HF. The antenna
structures are defined using optical lithography and reactive ion
etching.

The behavior of semiconductor antennas is determined by the doping
level, and the antenna geometry. The bowtie antenna is formed by two
monomers with a triangular shape with a base of
$\unit{100}{\micro\meter}$, a triangle height of
$\unit{300}{\micro\meter}$ and monomer height of
$\unit{1.5}{\micro\meter}$. The triangles have facing apexes,
separated by a gap of \unit{5}{\micro\meter}. As it is shown below,
this structure has a LSPR around \unit{0.4}{\tera\hertz}. A
schematic representation of the bowtie antenna and an optical
microscope image are shown in figures \ref{Fig02}(b) and
\ref{Fig02}(c), respectively. The cracks observed in the image are
cracks in the BCB layer. These cracks have a typical width of a
micron, being much smaller than the wavelength of THz radiation, and
do not influence the measurements.

\section{Simulations}\label{Sim}

The local field enhancement of the bowtie antenna and the
transmission properties of the conical waveguide are studied using a 3D finite
element method (FEM) in the frequency domain (COMSOL Multiphysics). For the
simulations, we use the antenna dimensions determined from the
optical image (figure \ref{Fig02}(c)). The dielectric constant of
n-doped Si at 0.4 THz is determined with the Drude model \cite{ashcroft1976,adachi2004} to be
$\epsilon=-2.1 \times 10^3 + i\cdot 1.7\times 10^{-3}$. For
simplicity, we consider the antenna to be in vacuum and illuminated
with a monochromatic continuous plane wave at normal incidence to
the plane of the antenna. The frequency of the simulations is
\unit{0.4}{\tera\hertz}. As it is shown in section~\ref{Exp},
the bowtie antenna has a LSPR around this frequency.

\begin{figure}%
\centering
\includegraphics[width=12 cm]{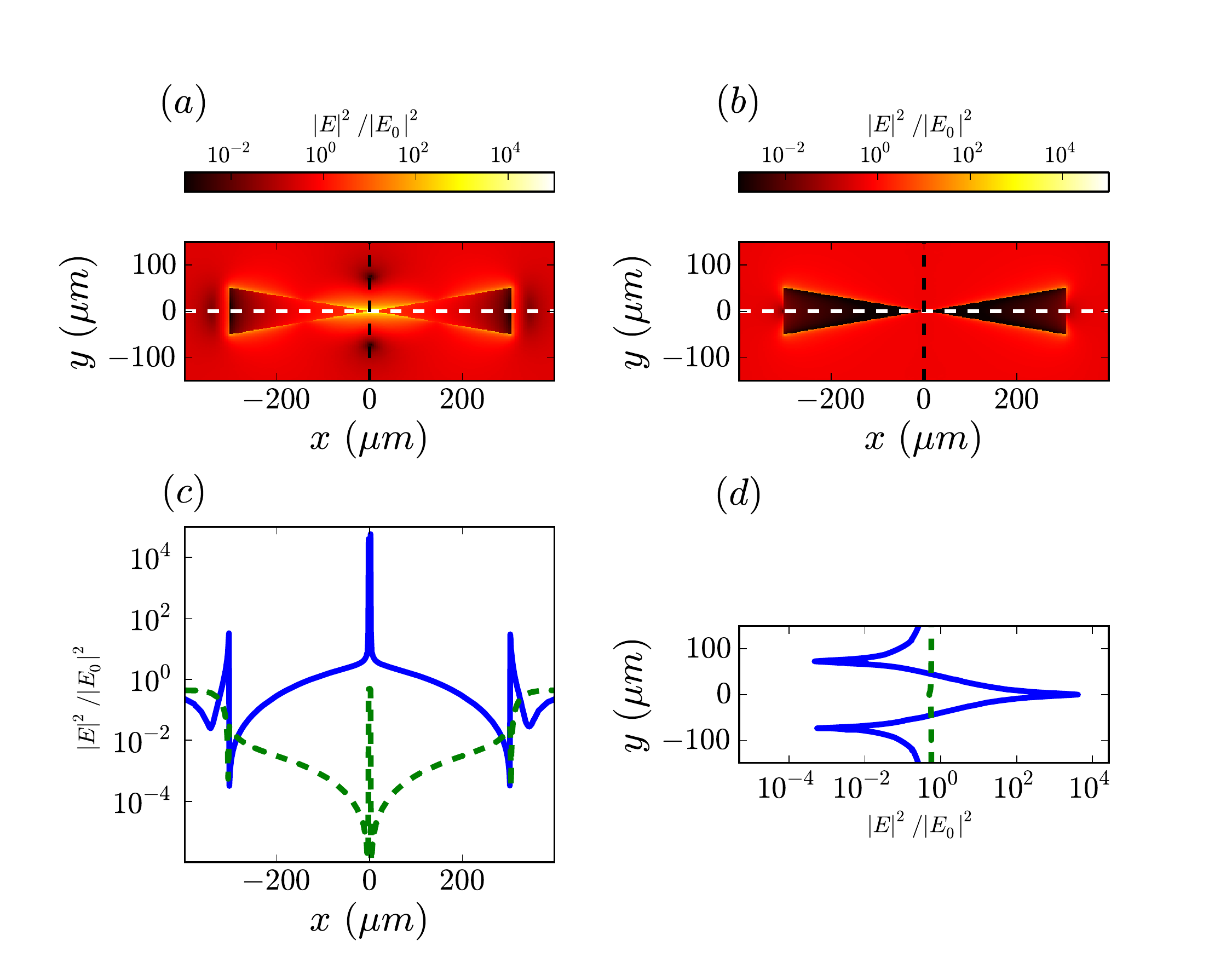}%
\caption{Simulations of the local field intensity enhancement close
to a doped silicon bowtie antenna formed by two triangular monomers.
The antenna is illuminated at normal incidence with a plane wave
having a frequency of \unit{0.4}{\tera\hertz}, i.e., with an
incident wavevector normal to the plane of the figure. (a) and (b)
show the local intensity enhancement $|E|^2/|E_0|^2$, where
$|E_0|^2$ is the incident field intensity, in the plane at the
middle height of the antenna for the incident THz radiation
polarized parallel and orthogonal to the long antenna axis (white
and black dashed lines), respectively. (c) Shows the intensity
enhancement along the long axis of the antenna for a polarization
parallel (blue-solid) and orthogonal (green-dashed)
to the long axis. (d) Shows the intensity enhancement for both polarizations along the short axis of the antenna.}%
\label{Fig03}%
\end{figure}

Figures \ref{Fig03}(a) and \ref{Fig03}(b) show the intensity enhancement of
the electric field in the plane at the middle height of the antenna,
normalized by the incident field. Figure \ref{Fig03}(a) displays
the intensity enhancement for a polarization of the incident wave along
the long axis of the antenna, while figure \ref{Fig03}(b) shows the
intensity enhancement for a polarization along the short axis. Cuts through
the intensity enhancement images along the long axis (horizontal-dashed lines)
are shown in Figure \ref{Fig03}(c), while cuts along
short axis (vertical-dashed lines) are shown in Figure
\ref{Fig03}(d). The blue-solid and green-dashed curves represent the intensity enhancement
for the polarization along the long axis and short axis,
respectively.

When driven along the long axis, the free electrons in the
semiconductor bowtie antenna resonate at the LSPR frequency. There
is a capacitive coupling of the two monomers across the gap. This
coupling, arising from the coulomb attraction of charges across the
short distance separating the two monomers, gives
rise to an intensity enhancement in the gap. The enhancement is
further increased by the lighting-rod effect that results from the
sharp tips forming the apex of the triangles. The simulations show an
intensity enhancement in the gap over 4 orders of magnitude. Note
that this enhancement might be different in a real sample due to
rounding of the antenna tips and the presence of the substrate.
Nevertheless, the simulations illustrate that resonant bowtie
antennas are capable of focusing and enhancing locally electromagnetic
fields in subwavelength volumes. For an incident polarization along
the short axis of the antenna there is no field enhancement
relative to the incident field at \unit{0.4}{\tera\hertz}. For this frequency and
polarization the antenna is not resonant with the incident THz wave.

\begin{figure}%
\centering
\includegraphics[width=12 cm]{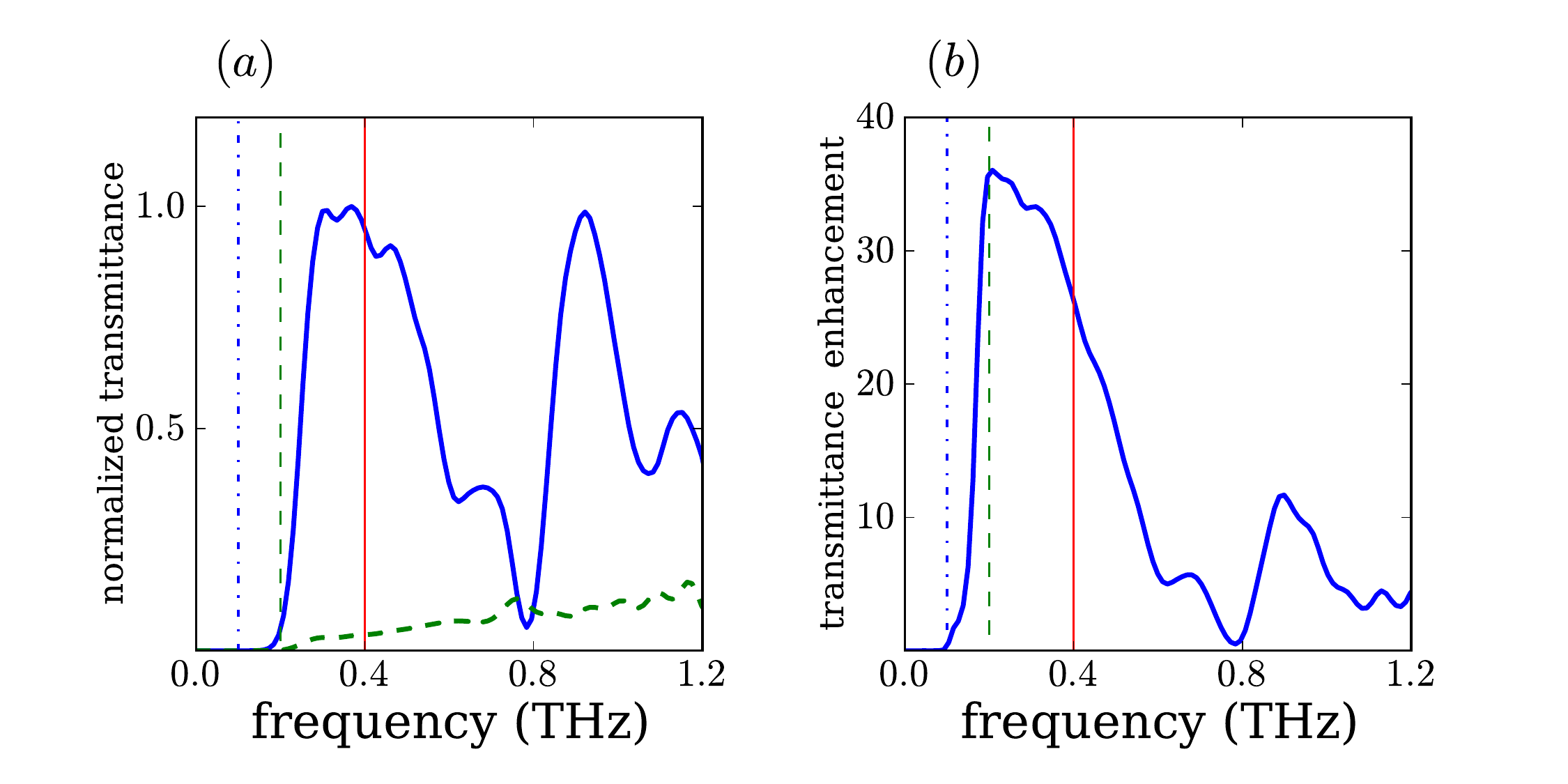}%
\caption{(a) Simulated transmittance spectra of a conically tapered
(blue-solid curve)
 and cylindrical (green-dashed curve) waveguide. (b) Transmitted intensity of
the conically tapered waveguide referenced by the cylindrical waveguide.
The vertical lines in both figures correspond to the frequencies at which
the field enhancements are shown in figure \ref{Fig05}.}%
\label{Fig04}%
\end{figure}

The transmission properties of the conically tapered and cylindrical waveguides
are simulated using FEM and finite-difference in time domain (FDTD,
 CST Microwave Studio) techniques. As sketched in figure
\ref{Fig01}, a linearly polarized plane wave travels from left to
right through the system. The dimensions of the waveguides used for
the simulations are those described in section~\ref{Sim}. The
aluminium forming the waveguides is simulated as a perfect electric
conductor. For the FDTD simulations a broadband Gaussian pulse is
sent through the system, and the intensity of the transmitted field
is monitored behind the waveguide, at a distance of
\unit{1.0}{\milli\meter} after the output exit. The transmission
spectra are obtained by Fourier transforming these fields. Figure
\ref{Fig04}(a) shows the transmittance spectrum, defined as the
transmitted intensity normalized by the incident intensity, of the
conically tapered waveguide with a blue-solid curve and of the
cylindrical waveguide with a green-dashed curve. Both simulations
have been normalized to the maximum transmittance of the conical
waveguide at \unit{0.37}{\tera\hertz}. At low frequencies the
transmittance vanishes. These frequencies are below the cutoff
frequency defined by the output aperture of the waveguides. For an
infinitely long cylindrical waveguide with a diameter of
\unit{0.6}{\milli\meter} the cutoff frequency is
\unit{0.3}{\tera\hertz}~\cite{marcuvitz1964}. Below cutoff the wave
entering the waveguide becomes evanescent. For a waveguide of finite
length, the evanescent field can still be transmitted for
frequencies just below cutoff.

Figure \ref{Fig04}(b) shows the transmittance enhancement, which is
defined as the transmittance through the conically tapered waveguide
normalized by the transmittance through the cylindrical waveguide.
The conically tapered waveguide enhances the transmitted field with
respect to the cylindrical waveguide. While the cylindrical
waveguide only transmits the fraction of the field incident onto the
opening and reflects the rest, the conically tapered waveguide
guides the incident field at the larger input aperture towards the
smaller output aperture. These fields travel a longer distance,
picking up an additional phase compared to the direct transmitted
fields. For the given waveguide dimensions this results in destructive
interference and a minimum in transmittance around
\unit{0.8}{\tera\hertz}. The entrance aperture of the conically
tapered waveguide has a radius of \unit{3}{\milli\meter}, which is
ten times the radius of the exit aperture. The 100 times larger energy
illuminating the conically tapered waveguide compared to the
cylindrical waveguide results in a maximum transmittance enhancement of around 35.

\begin{figure}%
\centering
\includegraphics[width=12 cm]{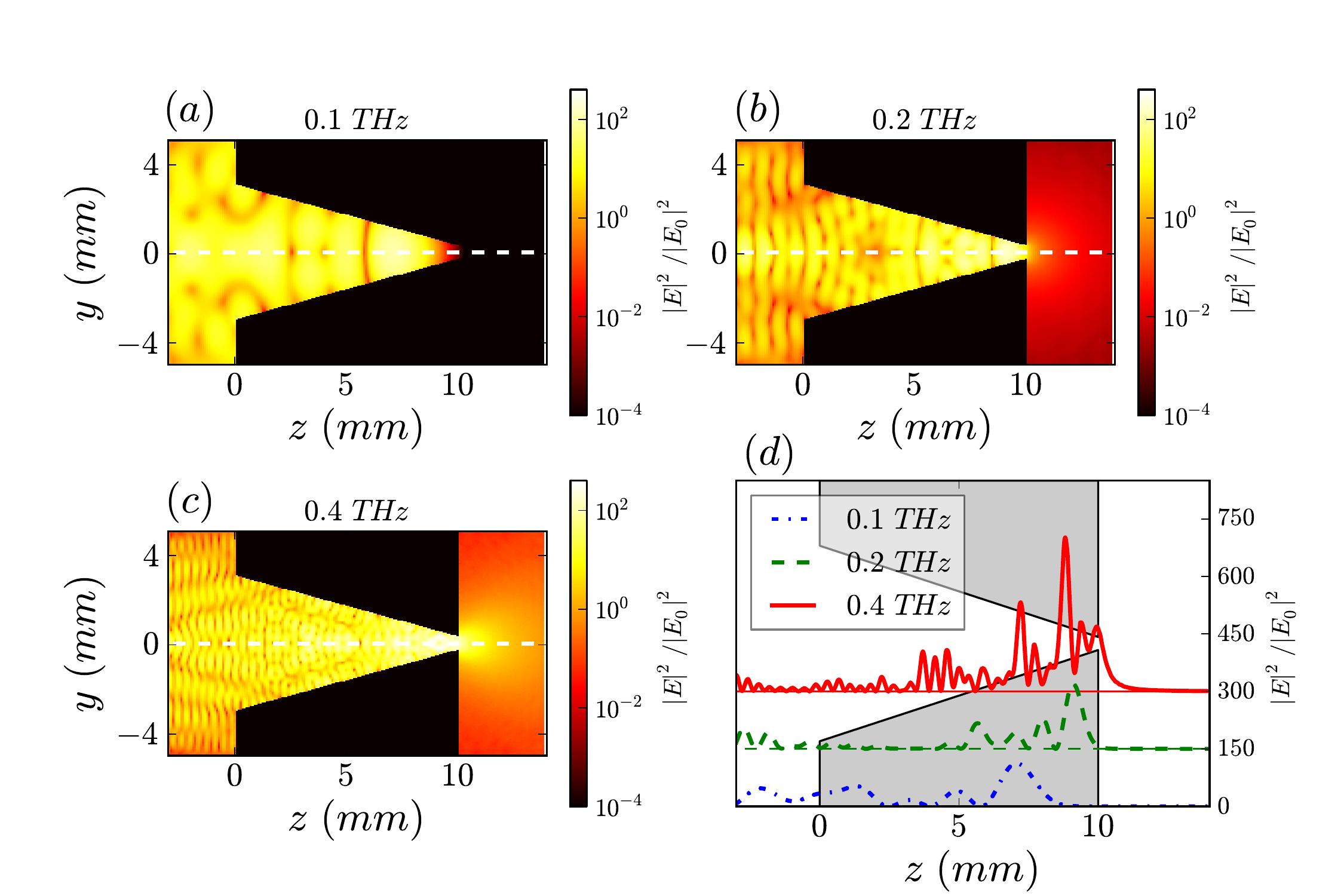}%
\caption{Simulations of the intensity enhancement of the THz
electromagnetic field propagating through a conically tapered waveguide at
various frequencies. A linearly y-polarized plane wave enters the
waveguide from the left and travels to the right.
2D cuts through the center of the waveguide are shown for frequencies
 of (a) \unit{0.1}{\tera\hertz}, (b) \unit{0.2}{\tera\hertz} and
(c) \unit{0.4}{\tera\hertz}. These frequencies correspond to
frequencies
 below, around and above the cutoff frequency of the waveguide, respectively.
Line cuts of the total field intensity enhancement through the center of the
 waveguide along the direction of propagation of the wave are shown in (d).
For clarity, the field intensities at 0.2 and 0.4 THz have been vertically displaced.}%
\label{Fig05}%
\end{figure}

Figure~\ref{Fig05} shows FEM simulations for the total field
intensity enhancement in the conically tapered waveguide. A
monochromatic and linearly y-polarized plane wave propagates from
left to right. The total field enhancement in the plane through the
center of the waveguide along the polarization direction -the yz
plane- is shown for frequencies below \ref{Fig05}(a), around
\ref{Fig05}(b) and above \ref{Fig05}(c) cutoff. Cuts through these
maps along the dashed white lines are shown in figure
\ref{Fig05}(d). For frequencies below cutoff there is an enhancement
of the field inside the waveguide, but the transmission is reduced
because the propagating field is fully reflected before the aperture
is reached. The interference pattern in the intensity enhancement is
the result of the interference of the incident field and this
reflection. For waves having frequencies close to the cutoff
frequency, there is a transition from the evanescent transmitted
field to propagating waves. Figure \ref{Fig05}(b) shows that
although most of the wave is reflected a small fraction is
transmitted. For frequencies well above cutoff, figure
\ref{Fig05}(c), the field propagates through the waveguide. This
condition is needed for a maximum enhancement of the transmission.
We note that even for this frequency, significant reflection takes
place and the interference pattern is formed. This reflection can be
minimized by reducing the tapering angle of the waveguide in order
to allow an adiabatic focusing of the incident THz field onto the
output aperture~\cite{rusina2008}.

\section{THz transmission and extinction measurements}\label{Exp}
The THz transmission experiments were carried out with a standard
THz-TDS setup \cite{dexheimer2008},
in which a Ti:Sapphire oscillator (Femtolasers, Fusion 20-800), providing
a \unit{75}{\mega\hertz} train of \unit{20}{\femto\second} pulses
around \unit{800}{\nano\meter}, is used to
generate broadband and linearly polarized THz pulses in a GaAs
photoconductive antenna. The detection is done with a ZnTe crystal
using the electro-optic effect. The THz beam has a Gaussian beam
profile with a full width at half maximum of \unit{1.7}{\milli\meter}. The waveguide is
placed in the THz beam, and the quartz substrate with the antenna
is clamped in front of the waveguide.

\begin{figure}%
\centering
\includegraphics[width=12 cm]{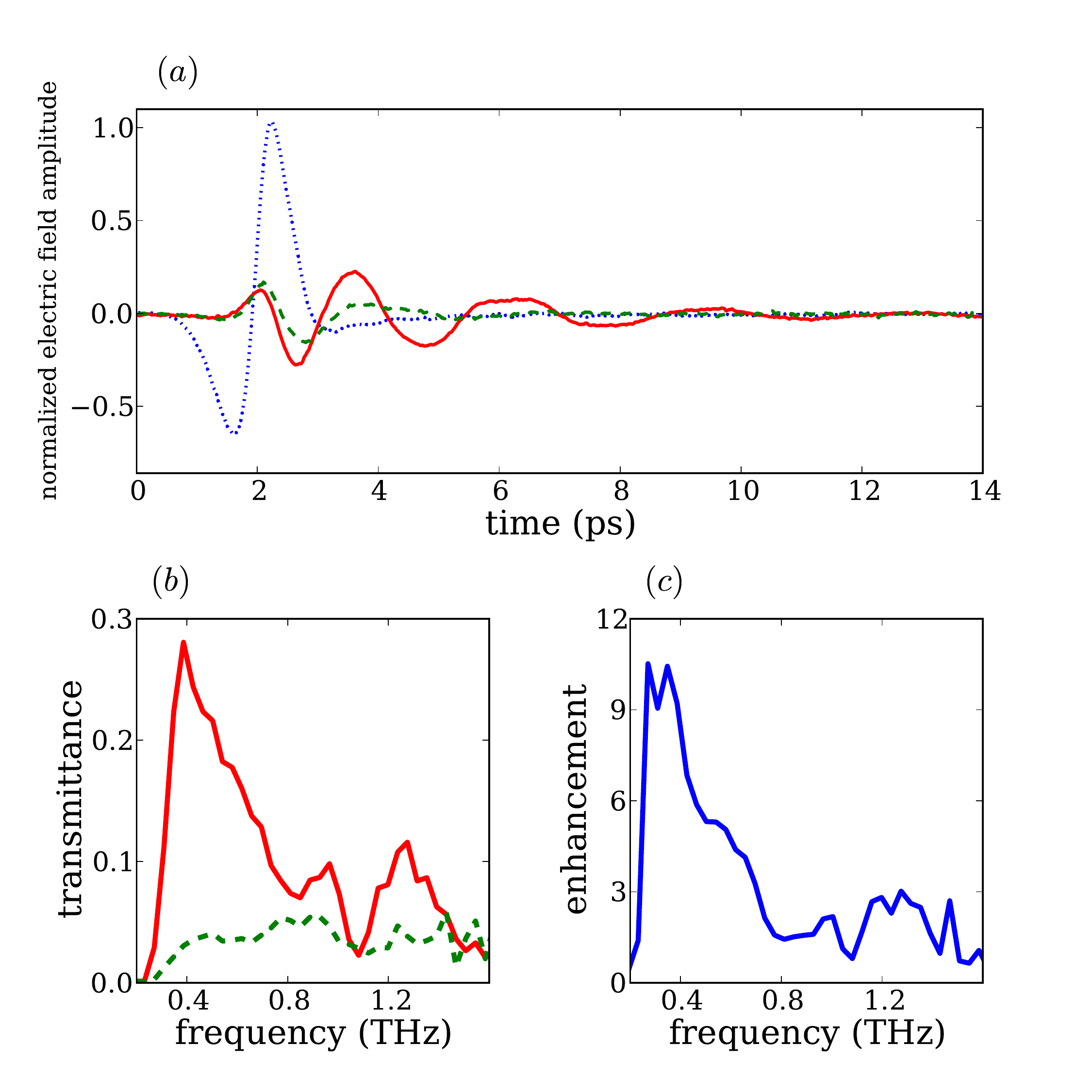}%
\caption{Experimental characterisation of the conically tapered and
cylindrical waveguides. (a) Time domain transmission measurements of
the reference (blue-dotted), conically tapered (red-solid) and cylindrical
(green-dashed) waveguide. (b) Fourier transform of the
transmitted intensity of the conically tapered (red-solid) and cylindrical
(green-dashed) waveguides
normalized by the reference. (c) Transmittance enhancement of the conically
tapered waveguide, defined as the transmittance through this waveguide normalized
 by the transmittance through the cylindrical waveguide.}%
\label{Fig06}%
\end{figure}

The experimental characterisation of the conically tapered and cylindrical
waveguides is shown in figure~\ref{Fig06}. Both waveguides are
measured in transmission and compared against each other and a reference
measurement taken without waveguide, i.e., the response function of the setup.
 The time domain signals (figure~\ref{Fig06}(a)) show that the presence of the
cylindrical waveguide severely reduces the transmitted signal, since
most of the incident amplitude is blocked by the waveguide. The
enhanced transmission of the conically tapered waveguide, relative to
the cylindrical waveguide, is visible as an increase in the THz
pulse dispersion. This is the contribution of the field that
illuminated the entrance of the waveguide at a larger radius, and
has picked up an additional phase before reaching the output
aperture.

The transmittance spectra for the conically tapered (red-solid
curve) and cylindrical (green-dashed curve) waveguides in figure
\ref{Fig06}(b) are obtained by Fourier transforming the time domain
signals to obtain the power spectrum, and are normalized against the
setup response, i.e., the power spectrum measured without any
waveguide. For both waveguides the transmittance vanishes at the
lowest frequencies due to cutoff. The transmittance remains below
0.05 for the cylindrical waveguide even at higher frequencies. This
reduced transmittance is due to the large area of the incident beam
that is blocked. In our setup, roughly 95 \% of the energy carried
by the THz beam is contained in an area of \unit{8}{\milli\meter^2},
while the area of the aperture of the cylindrical waveguide is
\unit{0.3}{\milli\meter^2}. This ratio of 1/25 matches the
experimental results. The transmittance is enhanced for the conical
waveguide, compared to the cylindrical one. Most of the
characteristics from the simulations shown in figure \ref{Fig04} are
reproduced by the measurements. The transmittance approaches 0.3 at
\unit{0.4}{\tera\hertz}, while the output aperture only encloses
around 5 \% of the area of the incident pulse. The reduced
transmittance around \unit{1.0}{\tera\hertz} is also consistent with
the simulations, and can be explained by the aforementioned
destructive interference in the wavefront. The blue shift of this
minimum in the measurements, compared to the simulations, can be
attributed to the approximation of plane wave illumination used for
the simulations. In the simulations, the complete input aperture is
illuminated by a plane wave, whereas in the experiments the pulse
has a Gaussian profile and it is slightly smaller than the input
aperture of the waveguide. This smaller size of the beam reduces the
effective height of the waveguide and introduces a blue-shift of the
interference features. Figure \ref{Fig06}(c) displays the
transmittance of the conically tapered waveguide normalized by the
cylindrical waveguide, showing the enhancement of the transmittance
of the conically tapered waveguide over the complete range of
\unit{0.1-1.4}{\tera\hertz}. The maximum intensity enhancement is
around one order of magnitude at \unit{0.4}{\tera\hertz}.

\begin{figure}%
\centering
\includegraphics[width=12 cm]{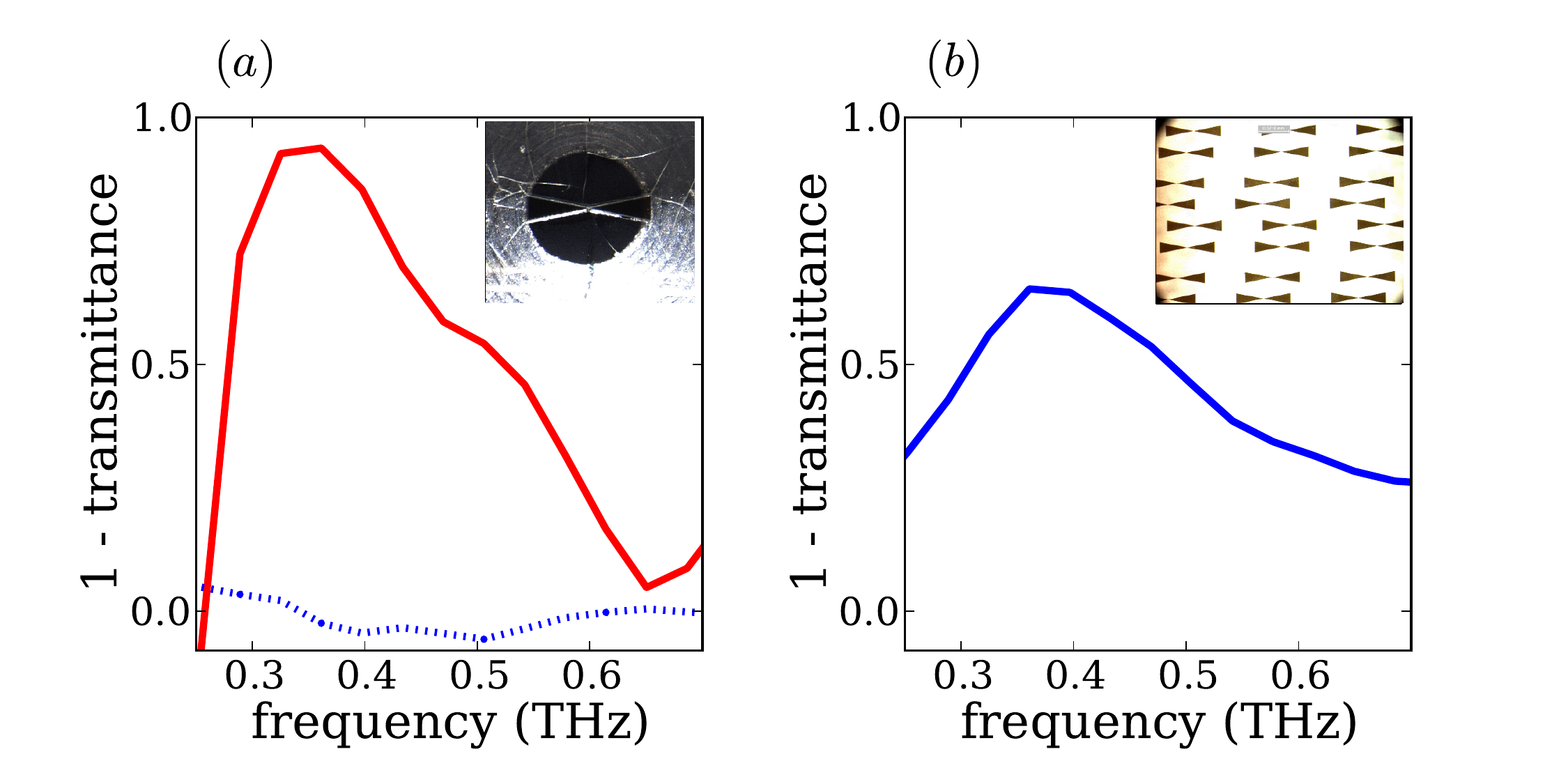}%
\caption{Extinction measurements of bowtie antennas, illuminated
with a polarization along the long axis and referenced against an
empty quartz substrate. (a) Extinction spectra of a single bowtie
antenna measured with the conically tapered waveguide (red-solid
curve) and without any waveguide (blue-dotted curve). In the inset
an optical microscope image of the antenna mounted in front of the
conically tapered waveguide is shown. (b) Extinction spectrum of a
random array of close packed antennas.
The inset contains an optical microscope image of the array.}%
\label{Fig07}%
\end{figure}

A single bowtie antenna is placed at the output aperture of the conically
tapered waveguide, as shown in the inset of figure \ref{Fig07}(a).
The measured extinction, which is defined as 1 minus the transmittance,
of this single antenna in front of the waveguide is shown in
figure~\ref{Fig07}(a) with the solid-red curve. The transmission measurements through the single
antenna are referenced to the transmission of the waveguide with an
empty quartz substrate at the output entrance, i.e., without the
antenna. A resonance is clearly resolved in this measurement with a maximum
extinction of 90 \% around \unit{0.4}{\tera\hertz}. This enhanced
extinction corresponds to the excitation of a LSPR in the plasmonic
antenna, which should lead to a large local field enhancement in the
bowtie gap. The blue-dashed curve in figure \ref{Fig07}(a) corresponds
to the extinction of the single antenna measured without the waveguide. The response is
practically flat and the LSPR cannot be
resolved.

To rule out any possible artifact in the measurements that could
lead to an extinction peak similar to our measurements, we have
confirmed the resonant response of the bowtie antenna by measuring
the extinction of a random array of similar antennas without the
conically tapered waveguide. An optical microscope image of the sample is
shown in the inset of figure~\ref{Fig07}(b). In this measurement,
the THz beam illuminates approximately 30 antennas. Therefore, the
extinction is enhanced in this sample by increasing the density of
antennas, rather than by concentrating the THz beam with the conical
waveguide. A similar extinction spectrum to the single antenna is
measured for the random array (figure \ref{Fig07}(b)). The extinction
reaches a maximum at 0.4 THz, with a resonant response that can be
attributed to the excitation of LSPRs.

The demonstration of the enhanced extinction by a single bowtie
antenna opens a range of possibilities for THz spectroscopy of
nanostructures or of molecules at low concentrations. For
example, the positioning of a nanostructure in the subwavelength gap
defined by the two monomers, in which the field is locally enhanced,
will enable to increase the THz field-matter interaction to a level at
which far-field spectroscopy of single nanostructures at THz
frequencies might be possible.

\section{Conclusion} \label{label006}
We have experimentally demonstrated that a conically tapered
waveguide can be used to funnel and enhance the THz intensity. This
intensity enhancement allows us to measure the extinction of a single
THz plasmonic bowtie antenna, which otherwise cannot be detected due to
the large background of unscattered radiation. The transmittance
properties are also investigated numerically finding an excellent
agreement with the measurements. The large localized field
enhancements that can be achieved by bowtie antennas in
subwavelength volumes, may open the possibility of using standard
far-field THz time domain spectrometers for the detection and
spectroscopy of single nanostructures.


\ack
This work is part of the research programme of the Foundation for
Fundamental Research on Matter (FOM), which is part of the
Netherlands Organisation for Scientific Research (NWO).
The work was partially supported by the European community's 7th
framework programme under grant agreement FP7-224189 (ULTRA project).

\end{document}